# Do Oceanic Convection and Clathrate Dissociation Drive Europa's Geysers?

Nicole C. Shibley[1] and Gregory Laughlin[2]
[1] Princeton University Center for Theoretical Science, Princeton, NJ, USA; nicole.shibley@princeton.edu
[2] Yale University Department of Astronomy, New Haven, CT, USA

## Abstract

Water vapor geysers on Europa have been inferred from observations made by the Galileo spacecraft, the Hubble Space Telescope, and the Keck Observatory. Unlike the water-rich geysers observed on Enceladus, Europa's geysers appear to be an intermittent phenomenon, and the dynamical mechanism permitting water to sporadically erupt through a kilometers-thick ice sheet is not well understood. Here we outline and explore the hypothesis that the Europan geysers are driven by $CO_2$ gas released by dissociation and depressurization of $CO_2$ clathrate hydrates initially sourced from the subsurface ocean. We show that $CO_2$ hydrates can become buoyant to the upper ice–water interface under plausible oceanic conditions, namely, if the temperature or salinity conditions of a density-stratified two-layer water column evolve to permit the onset of convection that generates a single mixed layer. To quantitatively describe the eruptions once the $CO_2$ has been released from the hydrate state, we extend a one-dimensional hydrodynamical model that draws from the literature on volcanic magma explosions on Earth. Our results indicate that for a sufficiently high concentration of exsolved $CO_2$, these eruptions develop vertical velocities of $\sim 700$ m s$^{-1}$. These high velocities permit the ejecta to reach heights of $\sim 200$ km above the Europan surface, thereby explaining the intermittent presence of water vapor at these high altitudes. Molecules ejected via this process will persist in the Europan atmosphere for a duration of about 10 minutes, limiting the timescale over which geyser activity above the Europan surface may be observable. Our proposed mechanism requires Europa's ice shell thickness to be $d \lesssim 10$ km.

*Unified Astronomy Thesaurus concepts:* Europa (2189); Ocean planets (1151); Volcanism (2174)

## 1. Introduction

Inferences from both photographs and magnetometer measurements obtained by NASA's Galileo orbiter generated a consensus that Europa harbors a subsurface ocean (e.g., Carr et al. 1998; Greeley et al. 1998; Kivelson et al. 2000). The ocean is presumed to be O(100) km deep and is covered by an ice cover that is likely O(10) km thick (e.g., Carr et al. 1998; Hussmann et al. 2002; Nimmo & Pappalardo 2016). The prospects inherent within a vast sea of geothermally warmed water have long drawn the attention and sparked the speculation of astrobiologists (e.g., Hand et al. 2009).

Possible evidence of liquid water in the near-surface environment was obtained when the Hubble Space Telescope detected the presence of hydrogen and oxygen in Europa's tenuous atmosphere (Hall et al. 1995; Roth et al. 2014). In isolated instances, water vapor geysers have been inferred from infrared spectra (Paganini et al. 2019), and hints of geysers have been found in the magnetic (Jia et al. 2018), chemical (Roth et al. 2014), and proton (Huybrighs et al. 2020) signatures of Europa's atmosphere. In particular, excesses of hydrogen and oxygen were detected with the Hubble Space Telescope at distances of $h \sim 200$ km above the surface of the $D \sim 3120$ km satellite (Roth et al. 2014). Treating these water-derived products as ballistic projectiles, a zeroth-order estimate for their ejection velocity from Europa's surface is then $v = \sqrt{2gh} \sim 700$ ms$^{-1}$, given Europa's $g = 1.31$ m s$^{-2}$ surface gravitational acceleration.

Venting of water is not unique to Europa in the outer solar system. The presence of geysers over the south pole of Enceladus, a smaller ice-covered moon of Saturn, has also been firmly established from absorption spectra (Hansen et al. 2006), dust measurements (Spahn et al. 2006), mass spectra (Waite et al. 2006), and photographic imaging (Porco et al. 2006) by NASA's Cassini spacecraft. The geysers of Enceladus are thought to be caused by ice sheet fissures opened via tidal stresses (e.g., Hedman et al. 2013), exposing water, and possibly other volatile species, to the near-vacuum of the satellite's exospheric environment (Hurford et al. 2007).

Unlike the consistent geysers of Enceladus (e.g., Hansen et al. 2006), the Europan geysers display only intermittent and transient observability. For example, in a series of observations by the Keck Observatory over 17 dates, only one date produced an emission spectra likely to be associated with a water vapor geyser (Paganini et al. 2019). This isolated event was interpreted to arise from the release of $\sim 2 \times 10^6$ kg of water into the thin Europan atmosphere (e.g., Paganini et al. 2019). Thus, the physical mechanism driving the ejection of water from Europa appears to be distinct from what is occurring on Enceladus (e.g., Kieffer et al. 2006; Nimmo et al. 2007).

The precise thickness of Europa's ice shell, through which the geysers may erupt, is not well known, and arguments are made for both "thick" and "thin" shells (e.g., Pappalardo et al. 1998; Greenberg et al. 1999; Hoppa et al. 1999; McKinnon 1999; Schenk 2002). For a thick ice shell, with an approximate depth of 10 km or more, the resulting temperature gradient between the Europan surface and the much warmer ocean may produce convective cells within the ice cover (e.g., McKinnon 1999; Schenk 2002). In contrast, for a thin ice shell, with a thickness of around 10 km or less (e.g., Schenk 2002), heat transport within the shell will be dominated by conduction (e.g., McKinnon 1999). The thickness, in turn, affects the ice shell's propensity to fracture; it is easier to generate cracks in a thin, brittle shell.

Multiple mechanisms for generating fractures have been suggested, with tidal stresses attracting particular note (e.g.,

Lee et al. 2005). Other ideas include the suggestion that overpressure generated between the liquid water and ice shell as liquid water freezes into ice may allow cracks to form in the ice shell (Manga & Wang 2007). Cracks alone, however, fall short in explaining the energetic geysering mechanism that appears to be occurring (e.g., Crawford & Stevenson 1988).

Several known terrestrial phenomena bear important similarities to the mechanisms that may be responsible for driving the cryovolcanic eruptions on Europa. Particularly apt analogies can be drawn from the dynamics of buoyancy-driven flows and pressure-mediated vapor exsolution that drive volcanic eruptions from Earth's interior. In a volcano, buoyant magma decompresses as it rises, releasing gas that further increases the magma's buoyancy (e.g., Woods 1995). Rising mixtures that approach the surface can then erupt from lithospheric vents. Similarly, kimberlite ($CO_2$-rich and occasionally diamondiferous melt sourced from the mantle) emplacement is generated by magma ascending through cracks in the Earth's lithosphere at high speeds. This process, which originates from depths of $\gtrsim 150$ km, permits self-perpetuating upward crack propagation through the lithosphere (e.g., Sparks 2013). When the pressure drops below a critical value, volatiles in the magma can exsolve in the crack tip, powering surface eruptions (Lister & Kerr 1991).

Similarly, the presence of volatile content dissolved in lake water may be responsible for limnic eruptions. Consider Lake Nyos in 1986. Here a lake in Cameroon, stratified in carbon dioxide, spontaneously released of order $V = 1$ km$^3$ of $CO_2$ gas (Kling et al. 1987). There are a variety of theories that describe the mechanisms precipitating this disaster, largely focusing on a buoyancy-driven flow propelled by the exsolution of carbon dioxide stored in the lake (e.g., Giggenbach 1990; Schmid et al. 2004).

Finally, methane clathrate hydrates (compounds in which molecules of methane are trapped within cage-like $H_2O$ crystal structures) produce crater-forming eruptions on Earth. These hydrates are stable at high pressure but will dissociate when pressure and temperature conditions permit (e.g., Buffett 2000), leading to a potentially explosive release of methane gas. For example, hydrate-driven explosions are thought to have occurred in permafrost (frozen soil; Vlasov et al. 2018) and undersea domes in the high-latitude regions (Andreassen et al. 2017).

We draw from these intensively studied terrestrial processes to propose an eruptive process on Europa driven by exsolved carbon dioxide. We propose that the gas derives from dissociated carbon dioxide clathrates. Our hypothesis builds on previous research that has speculated that cryovolcanic events driven by volatiles, and sometimes clathrates, may lead to geyser activity on both Europa and Enceladus (e.g., Stevenson 1982; Crawford & Stevenson 1988; Kieffer et al. 2006; Matson et al. 2012).

Specifically, we quantitatively develop a scenario in which buoyancy-driven oceanic flows permit $CO_2$ clathrates to migrate upward through cracks in Europa's ice shell to depths at which they become unstable. As discussed in detail in Section 3, for the process to work, the ice shell must be thin, with $d \lesssim 10$ km. At $P \sim 1$ MPa pressures, associated with $d \sim 1$ km of Europan ice overburden, the clathrates will dissociate, producing pressurized $CO_2$ gas. This generates an explosion of liquid water and $CO_2$ gas that fractures the ice and permits geysering with surface exit speeds of up to $\sim 700$ m s$^{-1}$, a velocity that is in line with the inferred presence of water vapor hundreds of kilometers above Europa's surface (e.g., Roth et al. 2014).

Our presentation is organized in the following manner. In the next section, we describe the conditions under which a two-layer Europan ocean with volatile-rich clathrates trapped at the interface between the layers can overturn, leading to clathrate buoyancy in the mixed column. Under sufficiently low-pressure conditions, the clathrates will dissociate, generating the eruptive process described above. In Section 3, we describe a one-dimensional hydrodynamical model for this eruptive process on Europa. The model indicates that the clathrate-eruption mechanism we propose can generate appropriate velocities to describe observations of water vapor above Europa's surface. Using this approach, we evaluate the timescales for atmospheric drawdown of carbon dioxide in Europa's atmosphere. In Section 4, we discuss how our hypothesis allows for future testable predictions from the Europa Clipper mission.

## 2. Oceanic Convection and Clathrate Buoyancy

### 2.1. A Two-layer Ocean Model

We start by outlining the conditions under which a two-layer Europan ocean can exist, as described in Zhu et al. (2017). Our proposed mechanism for intermittent geyser eruptions hinges on the density stratification, or lack thereof, between these two layers (Figure 1). Initially, the upper and lower ocean layers are separated by a density jump due to a freshwater cap generated by melting ice (Zhu et al. 2017). If conditions within the oceanic column change, however, and the density gradient between the layers decreases, then the entire column may overturn. In the event that a volatile substance (such as clathrated $CO_2$) is stored in the lower layer, this mixing has the potential to produce a water column in which volatiles at depth can become buoyant and propagate into a crack in the ice shell, eventually rising to a level where the pressure is low enough for dissociation to occur.

Europa's shell likely exhibits spatial variations in ice thickness (e.g., Ashkenazy et al. 2018). Ice may grow in the polar region of Europa, generating thick ice at high latitudes and thinner ice near the equatorial region. This ice thickness gradient generates a flux of ice from the region of thick ice to the region of thin ice. In a steady state, this pole-to-equator ice flow is balanced by ice melt in the equatorial regions. Here a cooler, fresher water layer (a "freshwater" cap), sourced from the melting ice, can rest atop a warmer and saltier water layer (Zhu et al. 2017).

The two-layer Europan water column is stably stratified (i.e., density increases with depth), provided the following condition is met:

$$\frac{\alpha \Delta T}{\beta \Delta S} < 1, \quad (1)$$

where $\alpha$ is the coefficient of thermal expansion, $\beta$ is the coefficient of haline contraction, $\Delta T = T_2 - T_1$ is the temperature jump across the upper and lower ocean layers, and $\Delta S = S_2 - S_1$ is the salinity jump across the layers. This condition dictates that the effect of salinity on density (a large effect at low temperatures) is larger than the effect of temperature on density, allowing a cold and fresh water layer to sit above a warm and salty water layer (Zhu et al. 2017).

Since any increase in ice thickness due to the ice flux must be balanced by melting of ice in a steady state, a relationship for the steady-state pole-to-equator flux of ice thickness ($F_h$) depends only on the diffusion of heat through the ice cover and

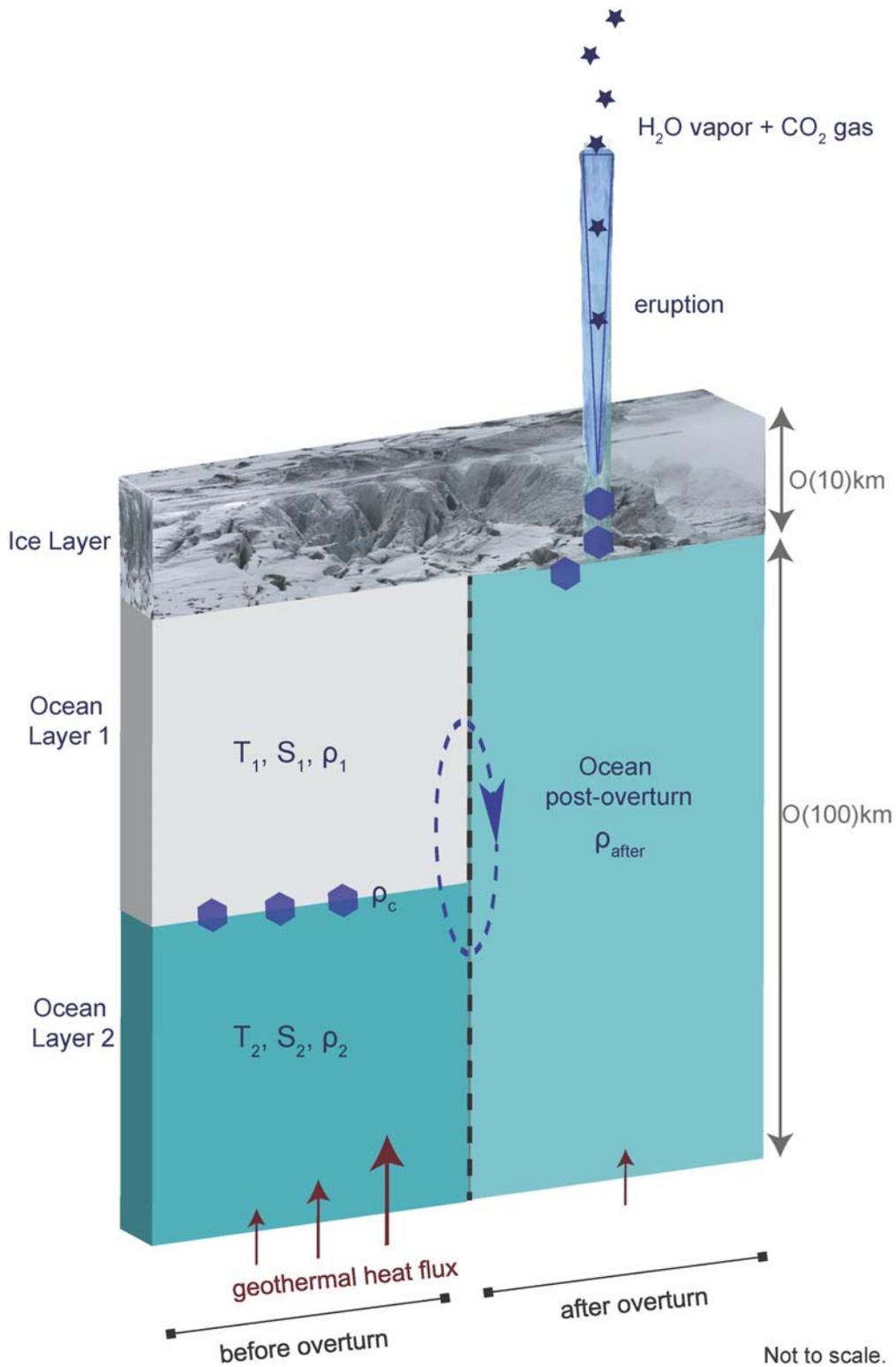

**Figure 1.** Schematic of Europa's ocean. The left-hand side shows the two-layer ocean preoverturn. An O(10) km thick ice layer overlies an O(100) km deep water column. The upper water layer has a temperature $T_1$, salinity $S_1$, and density $\rho_1$, and the bottom layer has a temperature $T_2$, salinity $S_2$, and density $\rho_2$, with carbon dioxide clathrates (hexagons) of $\rho_c$ massed at the interface between the two layers. An increase in the basal geothermal heat flux, for example, generates the conditions for an overturn. The right-hand side shows the resulting single-layer ocean with postoverturn density $\rho_{after}$. For specific density conditions, the clathrates will have $\rho_c < \rho_{after}$ and rise into any water-filled fissures present within the ice layer. At ~1 MPa (about 1 km) beneath Europa's surface, the clathrates will dissociate, creating $CO_2$ gas that triggers an eruption through the ice cover. At the surface, liquid water subject to near-vacuum conditions vaporizes, generating a transient high-altitude atmospheric geyser containing $H_2O$ vapor and $CO_2$ gas.

the ice-ocean heat flux. Here $F_h$ can be described as

$$F_h = \frac{\kappa_i \Delta T_{ep}}{2 h_0 L} + \frac{\Delta F_{ocn}}{2L}, \quad (2)$$

where $\kappa_i = 2$ W m$^{-1}$ K$^{-1}$ is the thermal conductivity of ice, $\Delta T_{ep} = 58$ K is the change in surface temperature between the equator and pole, $h_0 = 10$ km is the steady-state equatorial ice thickness, $L = 3.3 \times 10^8$ J m$^{-3}$ is the latent heat of fusion of ice, and $\Delta F_{ocn}$ is the change in ocean-ice heat flux between the equatorial and polar regions (see Zhu et al. 2017, for a description of these values).

This meridional ice flux then induces a saltwater flux in the opposing direction. This is because the corresponding ice melt in a steady state leads to a freshwater input to the subsurface ocean, while the corresponding ice growth leads to a saltwater input via brine rejection to the ocean below. The saltwater flux is written as $F_S = S_0(\rho_i/\rho)F_h$, where $F_S$ is the salt flux, $S_0$ is the salinity of the lower layer of the Europan ocean, $\rho_i$ is the density of ice, and $\rho$ is the density of liquid water (Zhu et al. 2017).

The salt flux can then be related to the salinity jump ($\Delta S$, small compared to $S_0$) across the upper and lower layers of the two-layer ocean. Salt is transferred from the lower layer to the upper layer by turbulent mixing. This is expressed as $cu^*\Delta S = F_S$, where $c$ is a parameter that depends on the ratio of interface stratification to shear, and $u^*$ is a turbulent velocity between the upper and lower ocean layers (Zhu et al. 2017).

The temperature jump between the upper and lower ocean layers is given by a balance between the geothermal heat flux ($F_b$) and ocean-ice heat flux. Assuming that the ocean water is not warming globally, the geothermal heat flux is related to the temperature jump across the upper and lower ocean layers as $F_b = \rho C_p cu^*\Delta T$, where $C_p$ is the specific heat capacity of the water. The stability condition (Equation (1)) can then be rewritten as

$$\frac{\alpha F_b}{\beta C_p \rho_i F_h S_0} < 1. \qquad (3)$$

This inequality quantifies the relative effects of geothermal heating, ice thickness flux, and background ocean salinity necessary to maintain appropriate density conditions for a stable column (see Zhu et al. 2017, for the full model description).

For the Europan ocean, which is thought to be laced with dissolved magnesium sulfate or sodium chloride, the minimum salinities consistent with the idealized two-layer ocean model of Zhu et al. (2017) are between 16 and 28 g kg$^{-1}$ (depending on whether the salt is NaCl or MgSO$_4$, respectively). They found that $S_0$ can be up to 200 g kg$^{-1}$ for a magnesium sulfate ocean and 100 g kg$^{-1}$ for a sodium chloride ocean, and that these values are within the bounds suggested by analyses of Europa's interaction with Jupiter's magnetic field (Hand & Chyba 2007; Zhu et al. 2017). However, it has also been suggested that Europa's salinity may be lower than the minimum values assumed here (e.g., Zolotov & Shock 2001; Hand & Chyba 2007; Steinbrügge et al. 2020). The exact value of salinity used for our hypothesis is not particularly critical, provided that it is high enough to permit establishment of a two-layer column and an intermediate clathrate density. Note that this does generally require a fairly salty Europan ocean, with a salinity of around 50 g kg$^{-1}$, in order to achieve the requisite ocean layer densities to bracket the clathrate density.

Further, our model for geysering hinges on the two-layer ocean being stratified by salinity rather than temperature. While it is still possible to attain a two-layer ocean even in a fresh scenario, the requirements for this are more rigid. In a pure freshwater ocean, a two-layer ocean configuration can be attained under sufficiently low pressure (e.g., Zhu et al. 2017), provided that the upper layer falls between 0°C and 4°C and the lower layer is at 4°C (e.g., Melosh et al. 2004), the temperature at which freshwater is densest. Then, if the lower layer is abruptly heated to above 4°C, the water will rise. However, if the lower layer is maintained at a temperature below 4°C, it must reach 4°C prior to any convective activity. Thus, while it may be possible to still achieve an overturn in a temperature-stratified freshwater ocean, this does not appear to be a particularly likely configuration given the large temperature differences required.

### 2.2. An Overturn

The Europan water column will overturn if the effect of temperature on density across the interface density jump is greater than the effect of salinity on density. To determine the exact conditions at which the water column will become unstable, we consider the marginal stability case, $\alpha\Delta T = \beta\Delta S$ (equivalent to $\alpha F_b = \beta C_p \rho_i F_h S_0$). The relevant dynamics arise from the flux terms, $F_b$ and $F_h$. We take all other values to be constant with fiducial values $\alpha = 7.7 \times 10^{-5}$ K$^{-1}$, $\beta = 7.7 \times 10^{-4}$ (g kg$^{-1}$)$^{-1}$, $C_p = 4000$ J kg$^{-1}$ K$^{-1}$, and $\rho_i = 920$ kg m$^{-3}$ and vary only $F_b$, $F_h$, and $S_0$ to reach conditions of marginal stability. These fiducial values are generally appropriate for an ocean containing either MgSO$_4$ or NaCl (Zhu et al. 2017).

The marginal stability case dictates the point beyond which a two-layer ocean would no longer be stably stratified and thus prone to forming a single layer with a uniform density. If $F_b/F_h > \beta C_p \rho_i S_0/\alpha$, the two-layer system will overturn. This means that if the geothermal heat flux at the base of the ocean increases sufficiently, the effect of temperature on density can cause the water column to be unstable. Similarly, if the ice thickness flux at the top of the layer decreases, the stabilizing effect of the freshwater cap will not be sufficient to prevent overturning.

Relatively minor changes in ice thickness flux or basal heating may locally destabilize the water column. Recall that the ice thickness flux can be related to the ocean-ice heat flux gradient by Equation (2) as in Zhu et al. (2017). For an ocean with two layers of equal thickness (50 km), a geothermal heat flux ($F_b$) of 0.01 W m$^{-2}$, a background salinity of $S_0 = 50$ g kg$^{-1}$, and an ocean-ice meridional heat flux ($\Delta F_{ocn}$) gradient of 0 W m$^{-2}$ corresponding to an ice thickness flux ($F_h$) of $1.76 \times 10^{-11}$ m s$^{-1}$, the water column is stably stratified (Zhu et al. 2017). However, for relatively minor changes in $F_b$ or $F_h$ (a factor of about 3 for $F_b = 0.01$ W m$^{-2}$), the stability condition is no longer satisfied. For example, at a geothermal heat flux of $F_b \gtrsim 0.03$ W m$^{-2}$ (and the parameters above), the water column will overturn. This value falls within the estimated range of geothermal heat flux, between 0.01 and 0.1 W m$^{-2}$ (Lowell & DuBose 2005; Zhu et al. 2017), and could be plausibly generated by spatially localized hydrothermal venting. Note that these increases in heat flux possibly generated by hydrothermal venting must be local phenomena. If the steady-state basal heat flux were higher, the salinity of the two-layer ocean would likely be tied to a different value (see Zhu et al. 2017).

An anomalous spatially localized ice thickness flux may also generate an instability. For example, the stability criterion is no longer satisfied for a flux $F_h < 5.4 \times 10^{-12}$ m s$^{-1}$, corresponding to a change in the meridional gradient in the ocean-ice heat flux of $-0.01$ W m$^{-2}$ (i.e., the ocean-ice heat flux at the poles is larger than the ocean-ice heat flux at the equator), per Equation (2). Under these conditions, the water column becomes unstable, and the upper and lower layers mix, yielding an intermediate density. Any substance with a density lower than this intermediate density, described above, will rise to the surface. This will provide the mechanism by which CO$_2$ clathrates, initially stored at depth in Europa's ocean, will dissociate under lower pressure, causing an explosion from Europa's ice shell.

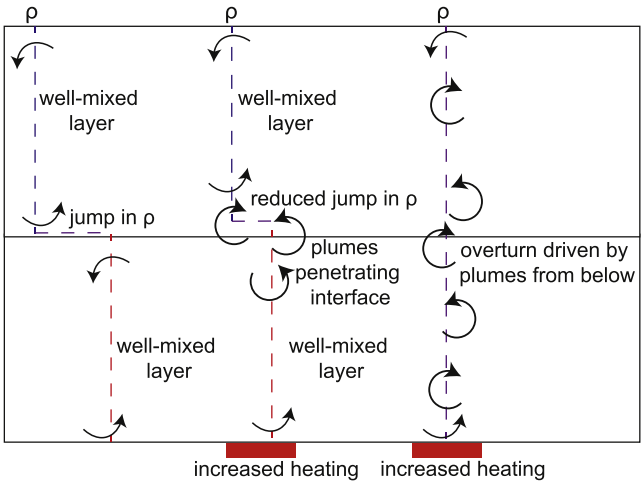

Figure 2. Schematic of the convective process leading to overturn of the water column. At first, two well-mixed layers (the cold freshwater layer and warmer, saltier layer) exist. Upon increased basal heating (for example), convective plumes from the lower layer penetrate the interface and mix up some fluid from the upper layer. In the final stage, plumes from the lower layer have fully penetrated into the upper layer, and both layers mix.

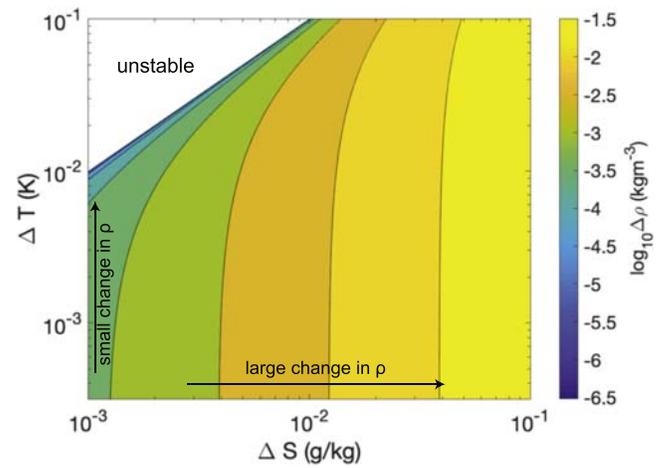

Figure 3. Change in density, $\Delta\rho$, between the top and bottom in a two-layer ocean as a function of temperature and salinity jumps $\Delta T$ and $\Delta S$ across the layer interface, based on the linear approximation for $\rho$. The representative density $\rho_0$ is taken to be 1060 kg m$^{-3}$, $\alpha = 7.7 \times 10^{-5}$ K$^{-1}$, and $\beta = 7.7 \times 10^{-4}$ (g kg$^{-1}$)$^{-1}$. Only $\Delta\rho > 0$ is shown. There is a restricted regime of temperature and salinity jumps where a two-layer ocean may be stable. Any perturbation from this would cause an overturn.

### 2.3. The Role of Carbon Dioxide Clathrates

Carbon dioxide clathrates, composites of water ice and $CO_2$ in a crystal structure, are hypothesized to exist in the Europan ocean (e.g., Prieto-Ballesteros et al. 2005; Bouquet et al. 2019). Their density, which depends on the temperature, the pressure, and the molar fraction of $CO_2$ in the crystal structure, is not fully known (Safi et al. 2017). For temperatures close to freezing and pressures of O(10–100) MPa (corresponding to depths of ∼10–100 km), $CO_2$-dominated clathrate densities are estimated to lie between ∼1040 and ∼1100 kg m$^{-3}$ (Prieto-Ballesteros et al. 2005; Bouquet et al. 2019). Remarkably, these clathrate densities are very similar to the typical densities of the Europan ocean inferred from hypothesized salinities (e.g., at 50 MPa, 273 K, and 20 g kg$^{-1}$, the ocean density would be 1039 kg m$^{-3}$; at 50 MPa, 273 K, and 100 g kg$^{-1}$, the ocean density would be 1100 kg m$^{-3}$). Our hypothesis for convectively driven geysers draws on this coincidental overlap in range and posits that a situation sometimes arises in which the clathrate density is bracketed by the densities of the upper and lower ocean layers. If this occurs, then the clathrates float at the interface of the two-layer ocean. After an overturn, any carbon dioxide clathrates with a density lower than the intermediate overturn density will rise toward the surface and eventually dissociate.

We can consider a situation where a two-layer fluid is heated from below; the heating will generate plumes emanating from the bottom layer that will eventually entrain fluid from the top layer. As a consequence, the lower layer progressively becomes less dense due to both the heating and the entrainment of the lighter overlying fluid. The lower layer will also thicken as it mixes in fluid from above (Figure 2). Simultaneously, the upper layer will increase its density slightly as it mixes with the fluid from below. As an end result, the entire upper layer can incorporate into the lower layer, yielding a single layer with an intermediate density (see, e.g., Davaille 1999, for a viscous case). As long as this intermediate density exceeds that of the clathrates, the clathrates will rise. Additionally, note that the inverse scenario will occur if the fluid is cooled from above (plumes will sink from above, mixing up the layer below), still leading to an intermediate density with buoyant clathrates.

We can illustrate this with the most simplistic scenario, where two layers with initial densities mix to create one layer of intermediate density. To fix ideas, we first assume that the upper ocean freshwater layer and the deep ocean layer are the same thickness, 50 km. The upper ocean density is $\rho_1 = \rho_1(T_1, S_1)$. The lower ocean layer density is $\rho_2 = \rho_2(T_2, S_2) = \rho_2(T_1 + \Delta T, S_1 + \Delta S)$, where $\Delta T = T_2 - T_1$, and $\Delta S = S_2 - S_1$. The density of the clathrates, $\rho_c$, follows $\rho_1 < \rho_c < \rho_2$, ensuring that they are sequestered between the two water layers. If the water column mixes, the full water column will have a density $\rho_{\text{after}} = \rho\left(\frac{T_1 + T_2}{2}, \frac{S_1 + S_2}{2}\right)$. If $\rho_c < \rho_{\text{after}}$, the clathrates will ascend in the water column.

To linear approximation, the density of the water can be written as $\rho = \rho_0(1 - \alpha T + \beta S)$, where $\rho_0$ is a reference density, $T$ is temperature, and $S$ is salinity. A change in density is thus $\delta\rho = \rho_0(-\alpha\delta T + \beta\delta S)$. We approximate the densities of the two layers in Europa's ocean before an overturn as $\rho_2 = \rho_0 + 2\delta\rho$ and $\rho_1 = \rho_0 - \delta\rho$, where $\rho_0$ is the background/reference density. We further take $\rho_c = \rho_0$. After a convective overturn, the density of the now one-layer system using this linear approximation will be $\rho_{\text{after}} = \rho_0 + 0.5\delta\rho$. Thus, a clathrate with density $\rho_c = \rho_0$ finds itself buoyant.

The temperature and salinity jumps at the interface of the upper and lower ocean layers under which a density jump can be maintained are rather restricted in scope (Zhu et al. 2017). For a background salinity of 50 g kg$^{-1}$ in an ocean composed of NaCl, a layer depth of 50 km, a prescribed 0.01 m s$^{-1}$ turbulent velocity between the upper and lower layer, and a geothermal heat source of 0.01 W m$^{-2}$, $\Delta T$ ranges from about $O(10^{-1})$ to $O(10^{-3.5})$ K, and $\Delta S$ ranges from about $O(10^{-1})$ to $O(10^{-3})$ g kg$^{-1}$, depending on the value of ocean-ice heat flux (see Figure 3(b) in Zhu et al. 2017). Taking $\alpha = 7.7 \times 10^{-5}$ K$^{-1}$ and $\beta = 7.7 \times 10^{-4}$ (g kg$^{-1}$)$^{-1}$ and selecting $\rho_0 = 1060$ kg m$^{-3}$ (to conform with clathrate density estimates and for an ocean at 50 g kg$^{-1}$) yields density jumps across the upper and lower layers ranging from $O(10^{-1.5})$ to $O(10^{-6.5})$ kg m$^{-3}$, shown in Figure 3. This indicates that the Europan ocean may be very weakly stratified and gives credence to the idea that a localized change in basal heat flux of a factor of 3 (see Section 2.2) may allow for an overturn.

In reality, a single oceanic overturn likely oversimplifies the dynamics within the Europan water column. More realistically, the configuration will be subject to penetrative convection. This mechanism describes an unstable fluid layer sitting below a stable layer, wherein plumes arising from the unstable layer eventually mix the fluid from the stable layer (e.g., Veronis 1963). The Europan setup of a freshwater layer sitting atop an ocean layer subject to destabilizing buoyancy forcing from hydrothermal vents seems apt for this process. We take the horizontal extent of such a convective cell to be the same order as the vertical scale of the convective cell, ultimately $\sim$100 km (although these cells may merge; see, e.g., Toppaladoddi & Wettlaufer 2018), giving an extent for the horizontal scale of penetrative convection in the Europan ocean as about 100 km. Taking a convective timescale (which describes how long it would take a fluid parcel to move from the lower layer to the upper layer) for this motion as $\sqrt{(l\rho_0)/(g\Delta\rho)}$ and using the range of values of $\Delta\rho$ from Figure 3 gives convective timescales between $\sim$14 and 4500 hr for high and low $\Delta\rho$, respectively. The lower end of this range is consistent with the inferred recurrence times for geyser eruptions ($\sim$1 every 30 hr) as seen by Paganini et al. (2019).

Moreover, a configuration in which colder, fresher water sits atop a warmer and saltier water layer may be conducive to double-diffusive convection. This diffusive–convective process, which is driven by the differential diffusivities between salt and heat, produces small-scale convective layers separated by sharp interfaces across which heat and salt are fluxed (e.g., Turner 1965; Shibley & Timmermans 2019). Past work has indicated that double-diffusive convection may be possible on Europa (Vance & Brown 2005), which would result in a vastly more complex problem. In particular, while double-diffusive convection would generally increase the density contrast between upper and lower ocean layers, depending on where the double diffusion is acting, it is also possible to homogenize portions of a water column through differential heat and salt fluxes. This may lead to smaller-scale vertical overturns with the potential to transport dissolved solute vertically, such as has been hypothesized for the case of Lake Nyos (e.g., Schmid et al. 2004).

## 3. A Model of the Geyser Mechanism

The convective process described in the last section will allow buoyant clathrates to propagate into fissures in Europa's ice shell. This motivates a hydrodynamical treatment to describe the time development of the process that erupts declathrated water and $CO_2$ on Europa. The framework is similar to the shock tube–inspired model of Turcotte et al. (1990), who modeled magma eruptions from Earth's lithosphere. If a crack or fissure exists in the ice sheet overlying Europa's ocean, then water will intrude up into the ice sheet to a height of $\sim$90% of the ice sheet thickness (based on the densities of ice and liquid water). Buoyant clathrates from the ocean will thus rise through the water intrusions following an oceanic overturn of the type described in the previous section. At a pressure of $\sim$1 MPa ($\sim$1 km below the ice sheet), the $CO_2$ clathrates will dissociate (Diamond & Akinfiev 2003), generating liquid water and carbon dioxide gas and paving the way for an explosion similar to the explosion that drives the cork from a champagne bottle (e.g., Liger-Belair et al. 2019). For an ice sheet of 10 km thickness, a fissure can be maintained by the denser liquid water up to 9 km. Above this point (at 1 km depth), not being maintained by adjacent water pressure, the ice walls will experience structural collapse. The clathrate dissociation at 1 km (1 MPa) provides a mechanism to remove this overlying ice cover, allowing for the geyser eruption. It is an interesting coincidence that the height to which ocean water will rise into an ice fissure on Europa with a 10 km thick ice sheet is the same depth at which clathrates will dissociate.

Simultaneously, such a clathrate-driven eruptive mechanism places an upper bound on the thickness of Europa's ice sheet. For an ice shell thickness $d \gtrsim 10$ km, water will not rise sufficiently far into a fissure to allow for an overlying ice cover of 1 km or less (see, e.g., Crawford & Stevenson 1988). (Water will only be able to propagate 18 km into a 20 km ice cover, for example.) Hence, if the mechanism we describe here occurs, it both signals and requires a constraint on Europa's ice thickness.

### 3.1. The Eruption

A carbon dioxide clathrate that rises from depth in the Europan ocean up to $\sim$1 MPa pressure will dissociate into water with dissolved carbon dioxide and excess exsolved carbon dioxide. The $CO_2$ solubility at 273 K and the 1 MPa pressure level appropriate to a 1 km depth in the Europan ice is $\sim$1.5 mol% (Diamond & Akinfiev 2003), corresponding to a saturated $CO_2$ mass fraction $\phi_0 \sim 3\%$–4%. Thus, if 1 kg of $CO_2$ clathrates with a composition of one molecule of $CO_2$ for $\sim$six molecules of $H_2O$ (Crawford & Stevenson 1988) are brought to 1 MPa, after dissociating, 0.02–0.03 kg of $CO_2$ will remain dissolved in the (saturated) water. Thus, 0.26–0.27 kg of excess $CO_2$ vapor per kilogram of clathrate brought to dissociation conditions will exist in the column. The amount of energy in the compressed $CO_2$ generated from bringing 1 kg of $CO_2$ clathrates to the dissociation pressure is sufficient to clear a conduit with at least an $A = 1$ cm$^2$ cross section in the overlying $d \sim 1$ km ice cover. As a consequence, in order to achieve an eruption of $2 \times 10^6$ kg of water vapor molecules (as in Paganini et al. 2019) from clathrate dissociation alone, $\sim 2.8 \times 10^6$ kg of clathrate must rise into the ice fissure, generating a postexplosion crater of $\sim$10 m in radius.

### 3.2. Governing Equations

After the initial explosion occurs, the water–carbon dioxide mixture will rapidly depressurize, and carbon dioxide will exsolve from the solution. The governing equations for a one-dimensional model of a fluid rising through a constant-area conduit are described here (Figure 4) and follow Turcotte et al. (1990). Variables (pressure $p$, vertical velocity $w$, density $\rho$, height $z$, time $t$, gas density $\rho_g$, mass fraction of gas $\phi$, and volume fraction of water $f$) are nondimensionalized, and hats denote nondimensional terms:

$$\hat{p} = \frac{p}{p_0}, \quad (4)$$

$$\hat{w} = w\left(\frac{\rho_{l0}}{p_0}\right)^{\frac{1}{2}}, \quad (5)$$

$$\hat{\rho} = \frac{\rho}{\rho_{l0}}, \quad (6)$$

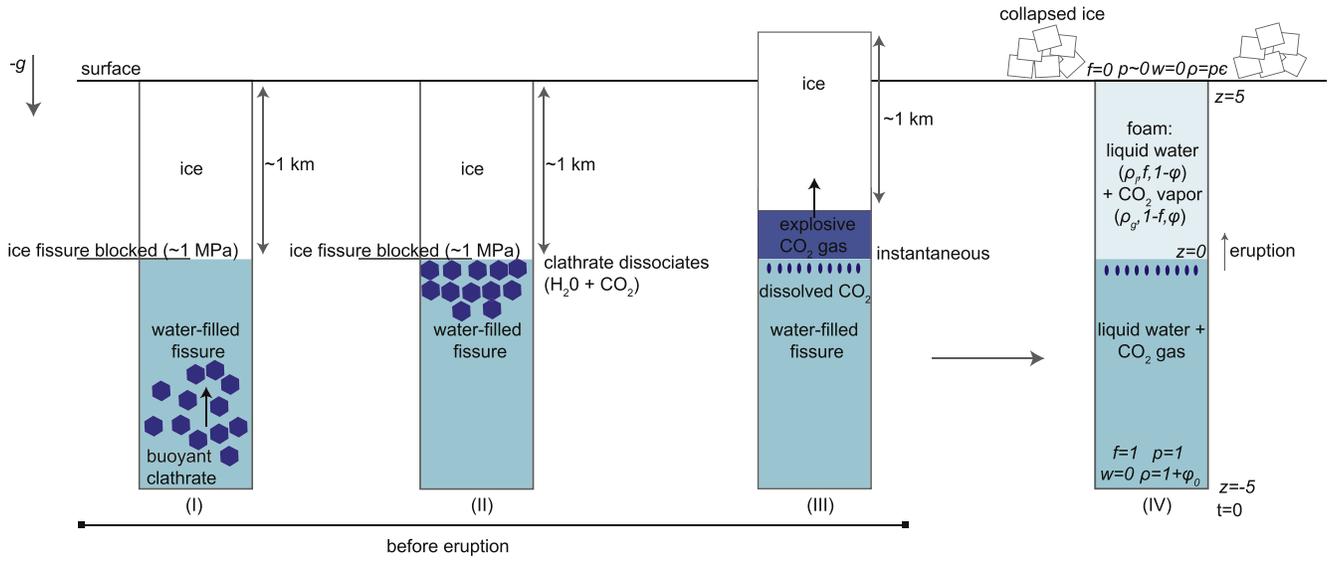

**Figure 4.** Schematic diagram showing an ice fissure at different stages in the eruptive process. In stage (I), buoyant clathrates, which are less dense than the fluid surrounding them, rise into a water-filled ice fissure. In stage (II), these clathrates sit at a pressure of 1 MPa, ~1 km deep in the ice shell. The clathrates then dissociate (at $z = 0$) under the low-pressure conditions, forming a mixture of $H_2O$ and dissolved $CO_2$, as well as excess $CO_2$ gas. In stage (III), the carbon dioxide gas is released explosively, blowing off the overlying 1 km of ice cover. In stage (IV), the $CO_2$ gas, previously dissolved in the water, comes out of solution, driving an eruptive geyser. A gas and liquid-water foam moves up the tube at high velocities, causing an eruption. The initial (boundary) conditions for pressure $p$, velocity $w$, density $\rho$, and volume fraction of liquid water $f$ are labeled. The dissolved vapor fraction at saturation, $\phi_0$, is between 0.03 and 0.04. All variables are nondimensionalized.

$$\hat{z} = \frac{z}{L}, \quad (7)$$

$$\hat{t} = t\left(\frac{p_0}{\rho_{l0}L^2}\right)^{\frac{1}{2}}, \quad (8)$$

$$\hat{\rho}_g = \frac{\rho_g RT_0}{p_0}, \quad (9)$$

$$\hat{\phi} = \frac{\phi}{\phi_0}, \quad (10)$$

$$\hat{f} = f, \quad (11)$$

where $p_0$ is the pressure at which the liquid contains a mass fraction $\phi_0$ of dissolved gas, $\rho_{l0}$ is the initial density of the pure liquid water prior to eruption, $L$ is a characteristic length scale, $R$ is the specific gas constant (defined as the ideal gas constant divided by the molar mass of the gas, which is 0.044 kg mol$^{-1}$ for carbon dioxide), $T_0$ is a reference temperature that we take to be the freezing temperature, and $\phi_0$ is the initial dissolved mass fraction of gas. Fiducial values for these quantities are $p_0 = 1$ MPa (equivalent to ~1 km below the Europan ice cover), $\rho_{l0} = 1000$ kg m$^{-3}$, $L = 200$ m, $R = 189$ J K$^{-1}$ kg$^{-1}$, and $T_0 = 0°C = 273$ K for a reference freezing temperature at the ice–water interface. The actual freezing temperature will be depressed slightly below $T_0$ due to the presence of salt. We take $\phi_0$ from 0.03 to 0.04 during different runs to represent the dissolved gas fraction of $CO_2$ after clathrate dissociation. Beyond this point, we drop the hats on nondimensional terms in the interest of clarity.

The following assumptions are invoked, similar to Turcotte et al. (1990):

1. The eruption can be modeled as a one-dimensional process.
2. The liquid water and the $CO_2$ gas move at the same speed.
3. The ice conduit area does not change in space or time.
4. Viscous stresses are small compared to the inertia of the fluid.
5. The temperature remains constant.

We extend the equations formulated by Turcotte et al. (1990) to include the influence of gravity. The resulting system incorporates conservation of mass,

$$\frac{\partial \rho}{\partial t} + \frac{\partial (\rho w)}{\partial z} = 0, \quad (12)$$

and conservation of momentum,

$$\rho\left(\frac{\partial w}{\partial t} + w\frac{\partial w}{\partial z}\right) = -\frac{\partial p}{\partial z} - \frac{1}{\text{EuFr}^2}\rho, \quad (13)$$

where the Euler number, $\text{Eu} = (p_0/\rho_{l0}w_0^2)$, measures the strength of pressure forces relative to inertial forces, and the Froude number, $\text{Fr} = w_0/\sqrt{gL}$, encapsulates the strength of inertial forces in comparison to gravitational forces. Additionally, the volume fraction, $f$, of liquid water obeys its own continuity equation:

$$\frac{\partial f}{\partial t} + \frac{\partial (fw)}{\partial z} = 0. \quad (14)$$

The nondimensionalized ideal gas law is given by

$$p = \rho_g. \quad (15)$$

An extension of Henry's law gives

$$\phi + p = 1, \quad (16)$$

and an equation for the density of the water–carbon dioxide mixture is

$$\rho = f[1 + \phi_0(1 - \phi)] + \epsilon(1 - f)\rho_g, \quad (17)$$

where $\epsilon = p_0/(\rho_{l0}RT_0)$.

Turcotte et al. (1990) neglected the gravitational term in the momentum equation and solved the resulting system. This gives the analytic solution $w = -\sqrt{\phi_0/(\epsilon + \epsilon\phi_0)} \ln p$, taking the condition that $w = 0$ for $p = 1$. We solve these equations (including the gravitational term) using a second-order Lax–Wendroff numerical scheme over 329 grid points (evenly spaced in increments of 0.03 nondimensional units). We confirm that there is a negligible difference between solutions with and without the gravitational term. For simplicity, we also neglect the gravitational term in the solutions described just below.

Noting that the six equations above are easily reduced to four, we adopt the following initial conditions for $w$, $\rho$, $f$, and $p$:

$$w(z, 0) = 0, \quad (18)$$
$$\rho(z, 0) = 1 + \phi_0 \quad z < 0, \quad (19)$$
$$\rho(z, 0) = p(z, 0)\epsilon \quad z \geqslant 0, \quad (20)$$
$$f(z, 0) = 1 \quad z < 0, \quad (21)$$
$$f(z, 0) = 0 \quad z \geqslant 0, \quad (22)$$
$$p(z, 0) = 1 \quad z < 0, \quad (23)$$
$$p(z, 0) = 10^{-13} \quad z \geqslant 0. \quad (24)$$

We apply Dirichlet boundary conditions, for which the quantities $w$, $\rho$, $f$, and $p$ at the upper and lower domain bounds are fixed to their initial values. Our numerical solution in the active region of the computational grid matches Turcotte et al.'s (1990) analytic solution.

We next examine how the eruption velocity of the carbon dioxide and liquid-water foam is affected by the saturated mass fraction of gas ($\phi_0$) and the time of the eruption at $p_0 = 1$ MPa below the ice shell (Figure 5). At early times (∼2 s) and $\phi_0 = 0.03$–0.04, consistent with the dissolved mass fraction of gas after clathrate dissociation, geyser eruption velocities can reach between 600 and 700 m s$^{-1}$ (Figure 5). As expected, larger dissolved mass fractions lead to larger vertical velocities. Vertical velocities of these magnitudes are large enough to be consistent with the existing inferences of water vapor above Europa's surface. Eruption velocities quickly decrease, reaching ∼130 m s$^{-1}$ roughly 10 s into the eruption.

Taken together, our results show that the inferences of Europan geyser activity can be explained by a hypothesis invoking a two-layer ocean bringing clathrates up to Europa's ice cover and a subsequent eruption generated by both the excess gas from clathrate dissociation and the dissolved gas in the water column under rapid depressurization.

Our theory allows for testable predictions. For example, an eruption of ∼$2 \times 10^6$ kg of water vapor (Paganini et al. 2019), arising from a solution with a 4% mass fraction of $CO_2$, would yield $8.3 \times 10^4$ kg of carbon dioxide; this should be observable from spectra. The $CO_2$ and $H_2O$ molecules released to the atmosphere will either undergo photodissociation or fall back down and potentially stick to the surface. These competing mechanisms allow for predictions of the observability of carbon dioxide and water molecules in Europa's atmosphere or on Europa's surface. We estimate an order-of-magnitude value for the rate of photodissociation of a $CO_2$ molecule using the solar flux (about 50 W m$^{-2}$; see Ashkenazy 2019), the absorption cross section of a $CO_2$ molecule (taken here as $10^{-25}$ m$^2$; see Schmidt et al. 2013), and the energy of a photon ($hc/\lambda = 1.1 \times 10^{-18}$ J, where $h = 6.6 \times 10^{-34}$ J s is Planck's constant, $c = 3 \times 10^8$ m s$^{-1}$ is the speed of light, and $\lambda = 180$ nm is a photon wavelength in the ultraviolet range). We assume that 5% of the solar flux is in the UV range. Then, an estimate for the rate of dissociation for a carbon dioxide molecule in Europa's atmosphere is $2.3 \times 10^{-7}$ s$^{-1}$. Further, a recent modeling study indicated that an estimated photolysis rate describing the dissociation of water into OH and H is $\approx 10^{-7}$ s$^{-1}$ (Li et al. 2020).

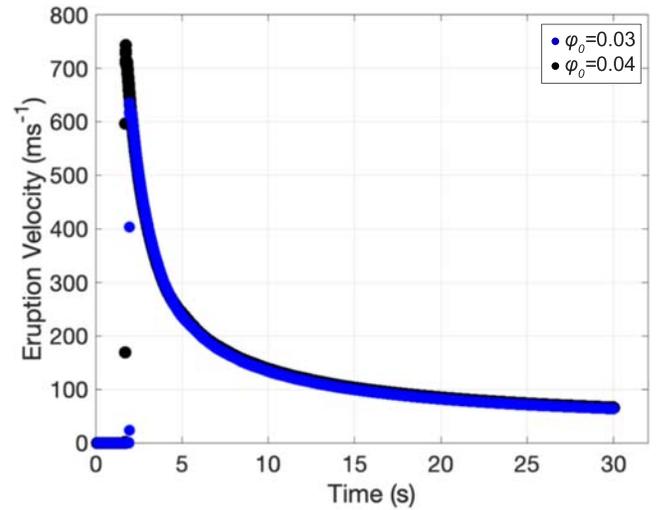

**Figure 5.** Eruption velocities at subsequent times for different initial content (3% in blue and 4% in black) of dissolved carbon dioxide. At early times (about 2 s after the ice is removed), eruption velocities are large enough to be consistent with inferences of water vapor found ∼200 km above Europa's surface.

If a molecule survives through its ballistic arc, it will stick to the Europan surface, provided that it loses enough energy on impact such that its resultant kinetic energy is smaller than the attractive potential between a water molecule at the surface and the fallen molecule. Using a simplified single-collision model (realistically, there will be multiple collisions) for an order-of-magnitude estimate, for a falling $CO_2$ molecule, this criterion gives

$$0.5 m_{CO_2} \left[ \frac{(\mu - 1)\left(v_{CO_2}^2 + \frac{2D}{m_{CO_2}}\right)^{0.5} + 2v_{H_2O}}{1 + \mu} \right]^2 \leqslant D, \quad (25)$$

where $m_{CO_2}$ is the mass of a carbon dioxide molecule, $\mu$ is the ratio of masses between the falling molecule and surface molecule, $m_{H_2O}$ is the mass of a water molecule, $v_{CO_2}$ is the velocity of the carbon dioxide molecule perpendicular to the surface, $v_{H_2O}$ is the velocity of a water molecule at the surface, and $D$ is the depth (energy) of the potential well (e.g., Hurkmans et al. 1976, who considered potassium on tungsten). The equivalent relationship for a falling water molecule becomes

$$0.5 m_{H_2O} v_{H_2O}^2 \leqslant D. \quad (26)$$

We take $v_{H_2O} = \sqrt{3 k_b T_s / m_{H_2O}}$ and estimate $D$ from the dipole-induced dipole interaction between a carbon dioxide molecule and a water molecule or the dipole–dipole interaction between water molecules, respectively. An order-of-magnitude estimate for $D$ gives $D = O(1)$ eV (using a distance of approach of ∼1 Å), while the left-hand side of the inequality is O(0.01–1) eV, indicating that a falling molecule (whether carbon dioxide or water) will likely adhere to Europa's icy surface.

Then, using a simplified ballistic representation, a molecule subject to a 1.3 m s$^{-2}$ surface gravitational acceleration will fall back down to Europa's surface in 554 s (about 9 minutes). For the small photolysis rates described above, this means that photodissociation of atmospheric carbon dioxide or water molecules will have a negligible effect on the number of geyser molecules observable above Europa's surface over a 9 minute period. If the eruptive mechanism that we have described occurs, then carbon dioxide and water vapor sourced from geysers will be observable in Europa's atmosphere on timescales of about 10 minutes.

With such a short timescale, one may expect to only intermittently observe the geysering phenomenon, consistent with the results of Paganini et al. (2019), who detected only one episode of increased water molecules in the Europan atmosphere out of 20 observations (totaling 1827 minutes of observations), implying a global eruption rate of roughly once every 30 hr. More precisely, out of these observations, only one instance of 148 minutes in length resulted in increased atmospheric water molecules (Paganini et al. 2019). Interestingly, this timescale is consistent with the lower values of the convective timescale (between ∼14 and 4500 hr) discussed in Section 2.3, providing a possible estimate of O($10^{-3}$) kg m$^{-3}$ on the density jump between the lower and upper layers of a stratified Europan ocean.

## 4. Discussion and Conclusions

The forthcoming launch of the Europa Clipper mission will allow for in situ investigation of the geysering phenomenon. In particular, the Europa Imaging System will provide a high-resolution (50 m or less) image of Europa's surface (Bayer et al. 2018); this may allow for observations of surface changes associated with the geyser process, which involves catastrophic disruption on horizontal scales of order ∼10 m and presumably subtler changes to surface shading over wider scales. Moreover, the synchronous measurements from the Clipper Ultraviolet Spectrograph and Mass Spectrometer may allow for chemical characterization of geyser molecules (Bayer et al. 2018), providing a further mechanism to test the hypothesis of a carbon dioxide–driven geyser.

At first glance, the possibility of geysers on Europa capable of transporting water vapor to ∼200 km above the satellite's surface presents a surprise. Europa's ice shell is estimated to be O(10) km thick, and this cold ice lid would not necessarily seem to lend itself to frequent eruptive activity. Indeed, in the terrestrial context, this picture is similar to that presented by the large subglacial water reservoir associated with Antarctica's Lake Vostok, which shows compelling analogies to the Europan structure (e.g., Wüest & Carmack 2000).

The novel aspect of our geyser hypothesis is that it connects a number of otherwise disparate phenomena. These include (1) the initiation of an unstable water column driven by changes in ocean heat fluxes or ice thickness fluxes, (2) the existence of faults or fissures (possibly driven by tidal stresses) in the overlying ice cover that permit the inflow of fluid, (3) a buoyancy-driven flow of carbon dioxide hydrates to any fissure in the overlying ice cover, and (4) a depressurization schema that can be fruitfully modeled with shock tube–like dynamics. The eruptive mechanism that we propose draws on a range of processes that have close analogs on Earth, such as the eruption of Lake Nyos, volcanic eruptions deriving from magma chambers, and methane hydrate–driven explosions from permafrost in the high latitudes. Yet while our hypothesis allows for water molecules to reach the vast heights necessary to be consistent with the inferences based on the Hubble observations (e.g., Roth et al. 2014), we must admit a certain degree of credulity to believe that the complex and interlinked chain of events occurs exactly as we describe in this paper. In fact, one wonders whether the eventual observations of the Europa Clipper mission will suggest that the geysers themselves were spurious inferences of water vapor, given the challenges necessary to ensure water vapor reaches the high altitudes required. If the process we suggest does occur, however, it provides detailed constraints on both the properties of Europa's subsurface ocean and the thickness of the ice shell itself.

This material is based upon work supported by the National Aeronautics and Space Administration through the NASA Astrobiology Institute under Cooperative Agreement Notice NNH13ZDA017C issued through the Science Mission Directorate. G.L. acknowledges support from the NASA Astrobiology Institute through a cooperative agreement between NASA Ames Research Center and Yale University. N.C.S. acknowledges support from the Princeton Center for Theoretical Science and the National Defense Science and Engineering Graduate Fellowship. G.L. and N.C.S. acknowledge valuable feedback from Dave Stevenson and an anonymous reviewer.

### ORCID iDs

Gregory Laughlin  https://orcid.org/0000-0002-3253-2621